\def \SAIT #1 #2 {{\em Mem.\ Soc.\ Astron.\ It.\/} {\bf #1}, #2}
\def \MESS #1 #2 {{\em The Messenger\/} {\bf #1}, #2}
\def \ASTRNACH #1 #2 {{\em Astron. Nach.\/} {\bf #1}, #2}
\def \AAP #1 #2 {{\em Astron. Astrophys.\/} {\bf #1}, #2}
\def \AAL #1 #2 {{\em Astron. Astrophys. Lett.\/} {\bf #1}, L#2}
\def \AAR #1 #2 {{\em Astron. Astrophys. Rev.\/} {\bf #1}, #2}
\def \AAS #1 #2 {{\em Astron. Astrophys. Suppl. Ser.\/} {\bf #1}, #2}
\def \AJ #1 #2 {{\em Astron. J.\/} {\bf #1}, #2}
\def \ANNREV #1 #2 {{\em Ann. Rev. Astron. Astrophys.\/} {\bf #1}, #2}
\def \APJ #1 #2 {{\em Astrophys. J.\/} {\bf #1}, #2}
\def \APJL #1 #2 {{\em Astrophys. J. Lett.\/} {\bf #1}, L#2}
\def \APJS #1 #2 {{\em Astrophys. J. Suppl.\/} {\bf #1}, #2}
\def \APSS #1 #2 {{\em Astrophys. Space Sci.\/} {\bf #1}, #2}
\def \ASR #1 #2 {{\em Adv. Space Res.\/} {\bf #1}, #2}
\def \BAIC #1 #2 {{\em Bull. Astron. Inst. Czechosl.\/} {\bf #1}, #2}
\def \JSQRT #1 #2 {{\em J. Quant. Spectrosc. Radiat. Transfer\/} {\bf #1}, #2}
\def \MN #1 #2 {{\em Mon. Not. R. Astr. Soc.\/} {\bf #1}, #2}
\def \MEM #1 #2 {{\em Mem. R. Astr. Soc.\/} {\bf #1}, #2}
\def \PLR #1 #2 {{\em Phys. Lett. Rev.\/} {\bf #1}, #2}
\def \PASJ #1 #2 {{\em Publ. Astron. Soc. Japan\/} {\bf #1}, #2}
\def \PASP #1 #2 {{\em Publ. Astr. Soc. Pacific\/} {\bf #1}, #2}
\def \NAT #1 #2 {{\em Nature\/} {\bf #1}, #2}
\def\bge{\begin{equation}}
\def\ede{\end{equation}}
\documentstyle[twoside]{memsait}
\input epsf.sty
\begin{opening}
\title{GRAVITATIONAL WAVES FROM COLLAPSING GLOBULAR CLUSTER SYSTEMS }
\author{ROBERTO CAPUZZO--DOLCETTA$^1$, PAOLO MIOCCHI$^2$}
\institute{$^1$Institute of Astronomy, University La Sapienza,
Roma, Italy\\
$^2$Dept. of Physics, University La Sapienza, Roma, Italy}
\date{} 
\end{opening}

\begin{document}

\oddpagefooter{}{}{} 
\evenpagefooter{}{}{} 
\bigskip

\begin{abstract}
 The evolution of globular cluster systems in some galaxies can be cause of merging of globulars in the very central regions.
This high stellar density favours the growth of a central nucleus
via swallowing of surrounding stars. The infall of stars into
a nuclear black hole is here shown to be, under certain conditions, not only source of electromagnetic radiation but also a significant source of gravitational waves.
\end{abstract}

\section{Introduction}
Many theoretical and observational arguments (Capuzzo--Dolcetta \& Vignola, 1997; Capuzzo--Dolcetta \& Tesseri, 1997; Capuzzo--Dolcetta, 1998; Capuzzo--Dolcetta \& Miocchi, 1998)
strongly support the hypothesis raised and discussed first in
Capuzzo--Dolcetta (1993) that the  AGN powering  can be,
at least in some galaxies, the energy subtracted to the gravitational field in form of stars belonging to a dense stellar
environment that are swallowed by the central massive black hole.
\par\noindent The dense stellar environment around the galactic nucleus is formed by globular clusters orbitally decayed to the central galactic regions due
to dynamical friction, and there partly merged and partly tidally
destroyed. The mass falling into the
nucleus goes both to enhance its mass and into radiation.
The process of matter infall corresponds to a non--radially simmetric nucleus accretion and so implies gravitational waves
emission, too.
Here we give a preliminary estimate of the number of  gravitational
bursts that the space--based interferometer
{\bf LISA} (at present under development by ESA) should detect.

\section{Gravitational waves from accreting galactic nuclei}

It is  known (see Davis \& Ruffini, 1971) that a point
mass $m$ in radial free fall into a black hole of mass $M$ emits an impulse of  gravitational
waves whose total energy is $E\sim 0.01(m/M)mc^2.$

It can be easily shown that the gravitational wave spectrum   peaks at the frequency
\bge
f_0\sim 0.05\frac{c^3}{GM}(z+1)\sim \frac{10^4}{M(\mbox{M}_\odot)}(z+1)
\ (\mbox{Hz}),
\label{freq}
\ede
and that the amplitude of the associated metric perturbation is
\bge
h\sim\frac{0.49}{d}\left(\frac{Gm}{c^2}\right),
\label{ampl}
\ede
where $d$ is the distance of the source  and $z$ its red--shift.

If the radius of the falling star, $R$, is  very small compared with the
wavelength of the gravitational waves emitted (of the order of the black hole
Schwarzchild's radius)
then the star can be actually treated as a point mass, and 
disruptive interference phenomena can be neglected
(see Haugan, Shapiro \& Wasserman, 1982 for more details).


 Flanagan \& Hughes  (1997) give 
the analytic approximation for the expected noise fluctuation curve $h_N(f)$ for the interferometer LISA,  showing
that the frequency range of acceptable gravitational waves
detection is $10^{-4} \div 10^{-1}$ Hz. A necessary condition for an impulse to be detected is:

\bge
S_N\equiv {h\over h_N(f_0)} >1. \label{snr}
\ede

where $S_N$, by definition, is the signal--to--noise ratio.
The condition on the star radius together with (1), (2)
and (3) reflect
into the following constraints involving the black hole mass and the distance of the source:

\begin{eqnarray}
M &>\! >& 6.7\cdot 10^5\frac{R^{9/8}}{m^{1/8}}~ (\mbox{M}_\odot),
\label{c1}\\
\nonumber \\
d &<& 2.4\cdot 10^{-14}\frac{m}{h_N(f_0)} ~(\mbox{pc}),\label{c2}
\end{eqnarray}
where $R$ and $m$ are expressed in solar units.
\par\noindent To give a global evaluation of the signal detectable it is
necessary, of course,  to integrate over all the possible
sources, i.e. to cumulate the contribution of all the stars falling into the
galactic central black holes in galaxies at any red--shift.

\section{Gravitational wave impulses from high density
galactic central regions}
As shown firstly by Capuzzo--Dolcetta (1993), dynamical friction and tidal
disruption are
effective mechanisms for globular cluster system (GCS) evolution in galaxies.
As a result, a globular cluster loses energy and angular momentum, approaching more
and more to the central regions where,
eventually, they can merge and grow a \lq super cluster\rq ~  and,
also, may lose their identity as a cluster due to a strong tidal interaction with a
compact central nucleus.
\par Capuzzo--Dolcetta (1993, 1998) computed a series of models 
of evolution of GCSs in a triaxial external potential,
taking into account collective effects (dynamical friction and
tidal interaction) to follow the mass loss to the galactic centre 
and its contribution to the galactic nucleus feeding and growth.
We do not deep here the details and characteristics of these models, but just say
that we refer to a model where: (i) the nucleus
is initially absent; (ii) the GCS system is composed by $1,000$ globular clusters
of the same mass, $M_{gc}=10^6$ M$_\odot$;
(iii) the clusters are 
moving on box orbits with a velocity dispersion about $300$ km/sec. The formation
and evolution of a central compact object
is followed in detail:
for what of interest here, we need just the quantities $\dot M(t)$ and $M(t)$,
i.e. the GCS mass loss and nucleus mass as functions of time.


\par We transformed the time dependences into red--shift
dependences assuming an Einstein--De Sitter universe and
noted that  an almost constant value of $M$ is reached at $z\simeq 0.35$, 
near to the peak of $\dot M$, for an assumed red--shift of galaxy formation
$z_f=5.5$.
This,
together with eq. (1), tells us that sources closer than $z\simeq 0.35$ (that is
1 Gpc) do not contribute to gravitational wave signal because they correspond to
too massive nuclei, $M> 10^8$ M$_\odot$ and generate impulses with $f_0$ outside 
the `optimal' $10^{-3} \div 10^{-2}$ Hz band. This 
implies a value of $m\geq 20$ M$_\odot$. 

For this reason we have to deal with massive stellar remnants, only.

\begin{figure}[htb]
\begin{center}
\leavevmode
\vskip 1 truecm
\epsfxsize 11 truecm
\epsffile{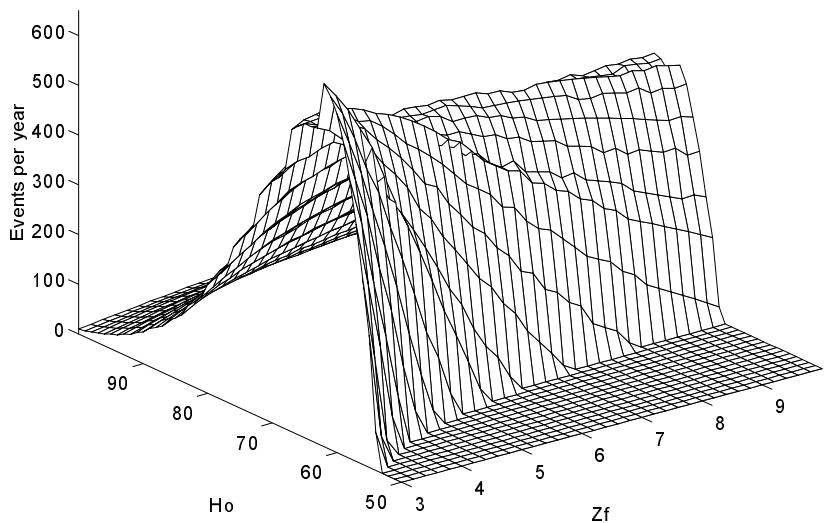}
\caption{Number of impulses per year that LISA should detect as a function
of $H_0$ and $z_f$ the red-shift of galaxy formation.\label{res}}
\end{center}
\end{figure}

We assume such remnants to be  compact objects with
$R<\!<1$ R$_\odot$ and $<m_r>= 30$ M$_\odot$.
The ratio of the number of these remnants to the total, $w$, is computed assuming 
a Salpeter's mass function
($\psi(m)\propto m^{-2.35}$). 
Given the number density of sources $n(z)=n_0(1+z)^3$, i.e.
the density at red-shift $z$ of galaxies presumed to have a central black hole,
and given the accretion rate $\dot M (z)$,
the number of events per unit time is just:
\begin{equation}
N=w\int_0^\infty {\dot M (z)\over {<m_r>}} \Theta (z)n(z)\Gamma(z)dz,
\end {equation}
where
\bge
\Theta (z)=\left\{
\begin{array}{rl}
1, & \mbox{\ if conditions (4) and (5) are satisfied} \\
0, & \mbox{\ otherwise,}
\end{array}
\right.
\ede
and $\Gamma(z)dz$ is the shell volume element.

According to our estimate, LISA should detect up to some
 hundreds events per year, depending on the value of the Hubble constant and on
the red-shift of galaxies formation $z_f$ (see Fig. 1).
This latter represents the origin of time by which
the AGN fueling starts to work.

\section{Conclusions}
The GCS evolution is responsible for AGN activity and,
moreover, the infall of compact stellar remnants
into the active central galactic black hole is source of gravitational waves.
These gravitational waves may be detect by the ESA 
space gravitational wave interferometer LISA which will be launched within year
2010. The number of detectable events per year is of the order of hundreds.

Our future aim is to improve the present treatment by considering also the
{\it inspiralling} of remnants towards the black hole and, by means of
a Fourier analysis of the impulses emitted, we wish to be able to
distinguish the kind of gravitational wave emission studied here from the
background of gravitational waves detected. 


\end{document}